# Symmetry breaking orbital anisotropy observed for detwinned Ba(Fe$_{1-x}$Co$_x$)$_2$As$_2$ above the spin density wave transition

Ming Yi[a,b], Donghui Lu[c], Jiun-Haw Chu[a,b], James G. Analytis[a,b], Adam P. Sorini[a], Alexander F. Kemper[a], Brian Moritz[a], Sung-Kwan Mo[d], Rob G. Moore[a], M. Hashimoto[a,b,d], Wei-Sheng Lee[a], Zahid Hussain[d], Thomas P. Devereaux[a,b], Ian R. Fisher[a,b], and Zhi-Xun Shen[a,b]*

[a]Stanford Institute for Materials and Energy Sciences, SLAC National Accelerator Laboratory, 2575 Sand Hill Road, Menlo Park, California 94025, USA

[b]Geballe Laboratory for Advanced Materials, Departments of Physics and Applied Physics, Stanford University, Stanford, California 94305, USA

[c]Stanford Synchrotron Radiation Lightsource, SLAC National Accelerator Laboratory, 2575 Sand Hill Road, Menlo Park, California 94025, USA

[d]Advanced Light Source, Lawrence Berkeley National Lab, Berkeley, California 94720, USA

* Corresponding author: zxshen@stanford.edu
Geballe Laboratory for Advanced Materials
McCullough Building 342
476 Lomita Mall
Stanford, CA 94305-4045
phone: (650) 725-0440
fax: (650) 725-5457

**Nematicity, defined as broken rotational symmetry, has recently been observed in competing phases proximate to the superconducting phase in the cuprate high temperature superconductors. Similarly, the new iron-based high temperature superconductors exhibit a tetragonal to orthorhombic structural transition (i.e. a broken $C_4$ symmetry) that either precedes or is coincident with a collinear spin density wave (SDW) transition in undoped parent compounds, and superconductivity arises when both transitions are suppressed via doping. Evidence for strong in-plane anisotropy in the SDW state in this family of compounds has been reported by neutron scattering, scanning tunneling microscopy, and transport measurements. Here we present an angle resolved photoemission spectroscopy study of detwinned single crystals of a representative family of electron-doped iron-arsenide superconductors, $Ba(Fe_{1-x}Co_x)_2As_2$ in the underdoped region. The crystals were detwinned via application of in-plane uniaxial stress, enabling measurements of single domain electronic structure in the orthorhombic state. At low temperatures, our results clearly demonstrate an in-plane electronic anisotropy characterized by a large energy splitting of two orthogonal bands with dominant $d_{xz}$ and $d_{yz}$ character, which is consistent with anisotropy observed by other probes. For compositions x>0, for which the structural transition ($T_S$) precedes the magnetic transition ($T_{SDW}$), an anisotropic splitting is observed to develop above $T_{SDW}$, indicating that it is specifically associated with $T_S$. For unstressed crystals, the band splitting is observed close to $T_S$, whereas for stressed crystals the splitting is observed to considerably higher temperatures, revealing the presence of a surprisingly large in-plane nematic susceptibility in the electronic structure.**

**Introduction**

Correlated electron systems owe their emergent phenomena to a complex array of competing electronic phases. Among these, a nematic phase is one where rotational symmetry is spontaneously broken without breaking translational symmetry (1-2). Two well established examples are found in certain quantum Hall states (3) and in the bilayer ruthenate $Sr_3Ru_2O_7$ (4), both of which exhibit a large transport anisotropy under the application of large magnetic fields, even though they seem to originate from apparently different physics. Recently, evidence of nematicity has also been reported in the pseudogap phase of cuprate high temperature superconductors (high $T_C$), in both $YBa_2Cu_3O_y$ (5) and $Bi_2Sr_2CaCu_2O_{8+\delta}$ (6). The proximity of the pseudogap phase to superconductivity raises the question of what role nematicity plays in relation to the mechanism of high $T_C$ superconductivity. Intriguingly, the newly discovered iron pnictide high $T_C$ superconductors also exhibit a nematic phase in the form of a tetragonal-to-orthorhombic structural transition that either precedes or accompanies the onset of long range antiferromagnetic order (7-8), both of which are suppressed with doping leading to superconductivity (9-11). Evidences of in-plane anisotropy have been reported by neutron scattering (12), scanning tunneling microscopy (13), and transport measurements (14-16). The physical origin of the structural transition has been discussed in terms of both spin fluctuations (17-21) and also orbital order (22-28). Here we present results of an ARPES study of underdoped $Ba(Fe_{1-x}Co_x)_2As_2$ that reveal an energy splitting of bands with principle $d_{xz}$ and $d_{yz}$ character that we show is associated with the structural transition. Although the splitting can be anticipated purely on symmetry grounds, the large magnitude (approximately 80 meV at 10 K for the parent compound) provides a quantitative test for theories of nematic order in this family of compounds. Moreover, we find that application of uniaxial stress causes the onset of the band splitting to occur well above the structural transition ($T_s$), indicating the presence of a large Ising nematic susceptibility.

In the orthorhombic phase, Ba(Fe$_{1-x}$Co$_x$)$_2$As$_2$ tends to form dense structural twins (29) which can easily obscure in-plane anisotropy. If the domains are large compared to the beam size, then information about the in-plane electronic anisotropy can be obtained by ARPES (30). Here, we apply an in-plane uniaxial stress to detwin the single crystals that we study, and hence avoid the problems of domain mixing. This also enables us to study the effect of uniaxial stress on the electronic structure *above* $T_S$.

**Results**

In Fig. 1 we compare the Fermi surfaces (FS) of twinned and detwinned BaFe$_2$As$_2$ ($T_S$ = $T_{SDW}$ = 138K) crystals in the SDW state measured under the same experimental conditions. Here we use the Brillouin zone (BZ) notations corresponding to the true crystallographic 2-Fe unit cell in which Γ-X is along the antiferromagnetic (AFM) crystal axis and Γ-Y is along the ferromagnetic (FM) crystal axis (Fig. 1a-c). In the twinned case (Fig. 1d), the nearly orthogonal domains mix signals from both Γ-X and Γ-Y directions, masking out any possible intrinsic differences between these directions and leading to complex FS topology and band dispersions. However, once detwinned, one clearly observes that the electronic structure along Γ-X (Fig. 1e) and Γ-Y (Fig. 1f) directions are different. Moreover, the anisotropy of the FS is further reinforced by the band dispersions measured along these two high symmetry directions (Fig. 1h-i), where the strong anisotropy of the band structure is not limited to near the Fermi level ($E_F$).

Next, we study the FS topology under different photon polarizations to reveal the underlying band character (Fig. 2a-c). For a multi-orbital system, variation of the orbital character around the FS can result in differed intensity patterns for different polarizations due to photoemission matrix element effects (see SI). The dramatic variation in intensity of the domain-resolved FS under different polarizations indicates that the FS of $BaFe_2As_2$ retains its strong multi-orbital character even in the SDW state, contrary to the claims of a recent laser-ARPES report on twinned crystals (31). Taking the FS pieces highlighted under different polarizations together, we arrive at the complete FS in the reconstructed SDW BZ (Fig. 2d), consisting of two hole pockets (α, β) and an electron pocket (γ) centered on Γ, surrounded by two bright spots (δ) along Γ-X and two bigger petal pockets (ε) along Γ-Y. A detailed $k_z$ dependence study on detwinned crystals shows that the qualitative anisotropy between $k_x$ and $k_y$ is robust for all $k_z$ values (see SI). As shown in Fig. 2a, the FS topology around the Y point is similar to the X point, as expected under BZ-folding after SDW reconstruction. However, $C_4$ rotational symmetry is broken, that is, the Γ-X and Γ-Y *directions*, which should be equivalent under $C_4$ rotational symmetry, are no longer the same. To ensure that this effect is not due to extrinsic photoemission matrix elements, we chose a polarization whose symmetry is equivalent for both Γ-X and Γ-Y directions, and yet still observe the same anisotropy (Fig. 2c). Hence, the observed difference unambiguously demonstrates that the x-y directional anisotropy reflects the intrinsic property of the electronic structure in the SDW state.

The most anisotropic features on the FS are the bright spots along Γ-X and petals along Γ-Y. As seen from band dispersions (Fig. 2e-f), these features are results of anti-crossing between hole and electron bands. Moreover, the bands cross at different energies in the two directions: close to $E_F$ along Γ-X inducing tiny Fermi pockets observed as bright spots (Fig. 2e), and 30meV below $E_F$ along Γ-Y (Fig. 2f) resulting in bigger electron pockets on the FS. These small Fermi pockets (0.2% and 1.4% of the paramagnetic (PM) BZ) are in good agreement with quantum oscillation reports (32). We note that such band crossings were discussed in early five-band model calculations where they were linked to Dirac cone features (33-34). The bright spots and petal-like FSs have been observed in earlier ARPES data on twinned $BaFe_2As_2$ (35-37) and $SrFe_2As_2$ (38). However, detwinned crystals here allow us to experimentally observe for the first time that the bright spots and petals reside along two orthogonal directions, manifesting the broken $C_4$ rotational symmetry of the electronic structure in the orthorhombic collinear SDW state.

To trace the origin of the anisotropy, we performed a temperature dependence study (Fig. 3a). The most anisotropic feature in the low temperature state is a pronounced hole-like dispersion near the X and Y points, which as shown evolves from the same hole-like dispersion in the PM state. Now considering the fact that polarization perpendicular to each high symmetry cut was used in Fig. 3a, only those bands with orbital character odd with respect to each cut direction can be observed. Hence for cuts along $\Gamma$-X, only bands of $d_{yz}$ and $d_{xy}$ orbitals can be observed; whereas for cuts along $\Gamma$-Y, only $d_{xz}$ and $d_{xy}$ orbitals can be seen. Moreover, $d_{xy}$ has very little intensity in the first BZ while $d_{yz}$ and $d_{xz}$ are much stronger in the corresponding geometries (see SI). Therefore, the prominent hole-like dispersion that we observe on these particular cuts must be $d_{yz}$ along $\Gamma$-X and $d_{xz}$ along $\Gamma$-Y, consistent with orbital assignments by local approximate local density approximation (LDA) (39).

Having identified the orbital characters of this anisotropic feature, we make the observation that for temperatures well above $T_{SDW}$ in the PM state of $BaFe_2As_2$, the $d_{yz}$ band along $\Gamma$-X and $d_{xz}$ band along $\Gamma$-Y are degenerate in energy, which is symmetric under the exchange of x and y axes, reflecting an inherent $C_4$ rotationally symmetric electronic structure as also confirmed by the measured FS (Fig. 3b). As temperature is lowered towards $T_{SDW}$, the degeneracy is lifted as the $d_{yz}$ band along Γ-X shifts up and crosses $E_F$ whereas the $d_{xz}$ band along Γ-Y shifts down, which are no longer equivalent under the exchange of x and y axes. This anisotropic band shift results in an unequal occupation of the $d_{yz}$ and $d_{xz}$ orbitals. The splitting between the originally degenerate $d_{yz}$ and $d_{xz}$ bands reaches ~60meV at 80K ($T_{SDW}$=138K) (Fig. 3a). Even more interestingly, a careful look at the detailed temperature evolution reveals that the anisotropic band shift persists above $T_{SDW}$, consistent with resistivity measurements reporting anisotropy developing above $T_{SDW}$ even for undoped compound held under uniaxial stress (15-16).

To further study this phenomenon, we measured underdoped compound Ba(Fe$_{0.975}$Co$_{0.025}$)$_2$As$_2$ detwinned by uniaxial stress, for which $T_S$ (99K) and $T_{SDW}$ (94.5K) are split. First, for this doping, we observe similar anisotropy in the electronic structure as that of the undoped parent compound (see SI). More importantly, the temperature dependence study (Fig. 3c) reveals that the anisotropic shift of the $d_{yz}$ and $d_{xz}$ bands is clearly present well above the long range magnetic ordering temperature, and for a temperature window even bigger than that of the undoped parent compound, consistent with transport results (15). To quantitatively illustrate the band shift with temperature, we plot the energy positions of the $d_{yz}$ and $d_{xz}$ bands at the same representative momentum position along the Γ-X and Γ-Y high symmetry lines (for analysis details see SI), revealing a trend that is reminiscent of the anisotropy observed above $T_{SDW}$ in bulk sensitive resistivity measurements (Fig. 3e). Hence, our ARPES study here reveals that the intriguing anisotropy seen in the resistivity is associated with a large $C_4$ symmetry breaking modification of the electronic structure that develops almost fully above the long range magnetic order.

To gauge the effect of the uniaxial stress, we also measured unstressed twinned samples for comparison, for which the energy separation between the $d_{yz}$ and $d_{xz}$ bands can be measured as they appear on the same cut due to twin domain mixing (Fig. 3d). We observe that below $T_{SDW}$ the splitting in the unstressed sample has the same magnitude within experimental uncertainty as that found for the stressed crystals. However, whereas the splitting in the stressed samples persists well above $T_S$, it decreases rapidly starting around $T_{SDW}$ and diminishes slightly above $T_S$ for the unstressed samples (Fig. 3e). That is, for an unstressed crystal, the electronic structure respects $C_4$ symmetry well above $T_S$. Approaching $T_S$, rotational symmetry is broken as an anisotropic band shift rapidly develops between the $d_{yz}$ and $d_{xz}$ orbitals, leading to a symmetry broken electronic structure in the orthorhombic state (Fig. 4). In the case of a stressed crystal, the extended temperature window above $T_S$ in which anisotropy persists can be understood as arising from the small uniaxial stress in the sample which acts as a symmetry breaking field for the structural phase transition, smearing out $T_S$ into a crossover temperature while leaving $T_{SDW}$ well-defined and little affected with increasing pressure (15). The extended temperature window above $T_{SDW}$ in which anisotropy persists induced by the uniaxial stress reveals a large intrinsic electronic nematic susceptibility. We note that while a lattice distortion exists below $T_S$, it is too small to account for the large magnitude of the energy splitting we observe in the orthorhombic PM state, as a non-magnetic band calculation accounting for the lattice distortion shows only a splitting of 10meV (see SI), well below the observed value of up to 80meV.

We also carried out a more detailed doping dependence study of the anisotropic energy splitting between the $d_{yz}$ and $d_{xz}$ bands at a temperature (10K) well below $T_{SDW}$ on both twinned and detwinned crystals in the underdoped region (see SI), which confirms that the full magnitude of the splitting is little affected by the uniaxial stress. Moreover, the magnitude of the splitting decreases monotonically with doping (Fig. 3f) — a trend consistent with the suppression of the magnetic ordering temperature and structural distortion (40), as well as optical conductivity measurement on detwinned crystals showing smaller dichroism for doped samples compared to the parent compound (41). This composition dependence contrasts with transport measurements, for which the maximum in-plane resistivity anisotropy is observed not for zero doping, but dopings closer to the superconducting dome (15).

**Discussion**

Our findings have several important implications for theories of the electronic structure of the iron pnictides. First, we observe that a large rotational symmetry breaking modification of the electronic structure occurs from the tetragonal PM state to the orthorhombic SDW state, providing a microscopic basis for the many signs of in-plane anisotropy reported in literature (12-16, 41). Furthermore, this broken symmetry is also revealed here to be manifested in an unbalanced occupation in the $d_{yz}$ and $d_{xz}$ orbitals, which develops almost fully at $T_{SDW}$. This observation alone does not distinguish the origin of the structural phase transition. The large amplitude of the band splitting draws attention to the possibility that the orbital degree of freedom might play an important role. In this context, it is worth noting that while it is experimentally difficult to calculate the orbital anisotropy over the whole BZ, a simple estimate based on a modified band structure calculation incorporating experimentally observed orbital splitting shows the orbital anisotropy over the entire FS to be on the order of only 10-20% (see SI), which is too small to be associated with a simple Kugel-Khomskii type orbital ordering (42). Finally, the electronic anisotropy seen in the orbital degree of freedom is observed to onset at a temperature well above the structural transition in stressed crystals. This temperature window, which indicates the involvement of fluctuations, appears to be bigger for doping levels closer to the superconducting dome, a trend also seen in transport measurements, suggesting that fluctuation effects, possibly of orbital origin, may play an important role in high temperature superconductivity in the iron pnictides.

**Materials and Methods**

High quality single crystals of $Ba(Fe_{1-x}Co_x)_2As_2$ were grown using the self flux method (10). Detwinned single crystals of $Ba(Fe_{1-x}Co_x)_2As_2$ were obtained using a modified version of the mechanical device reported by Chu *et al* (15), where uniaxial stress was applied at room temperature, and stressed crystals cooled down to measurement temperature before cleaving in-situ. ARPES measurements were carried out at both beamline 5-4 of the Stanford Synchrotron Radiation Lightsource and beamline 10.0.1 of the Advanced Light Source using SCIENTA R4000 electron analyzers. The total energy resolution was set to 15 meV or better and the angular resolution was 0.3˚. Single crystals were cleaved in situ at 10 K for low temperature measurements, 80K (60K) for temperature dependence measurements on undoped (doped) compound, and 150 K for high temperature measurements. All measurements were done in ultra high vacuum chambers with a base pressure lower than $4\times10^{-11}$ torr.

We note that our detwinning method does not affect the intrinsic electronic structure of the crystal since a mixture of the FS and band dispersions along the two directions on detwinned crystal reproduces those observed on the twinned crystal, as expected under domain mixing (Fig. 1). Furthermore, the high degree of detwinning is evident in contrasting band dispersions along the two orthogonal directions. For example, the bands highlighted by the arrow along Γ-X (Fig. 1h) are almost indiscernible along Γ-Y on a highly detwinned sample (Fig. 1i), whereas for a partially detwinned crystal, faint traces of these bands would be seen along Γ-Y due to mixing of Γ-X and Γ-Y.

**Acknowledgements**

The authors are grateful for helpful discussions with C.-C. Chen, R.-H. He, I. I. Mazin, Y. Ran, D. J. Singh, F. Wang, D.-H. Lee, J. P. Hu, Z. Y. Lu. ARPES experiments were performed at the Stanford Synchrotron Radiation Lightsource and the Advanced Light Source, which are both operated by the Office of Basic Energy Science, U.S. Department of Energy. The Stanford work is supported by DOE Office of Basic Energy Science, Division of Materials Science and Engineering, under contract DE-AC02-76SF00515. MY thanks the NSF Graduate Research Fellowship Program for financial support.

**Figure Legends**

**Fig. 1: Anisotropy of electronic structure observed on detwinned BaFe$_2$As$_2$**

(a) Schematic of unit cell of BaFe$_2$As$_2$ in the orthorhombic SDW state where spins are FM along the shorter *b* axis and AFM along the longer *a* axis. (b) Schematic of 3D BZ for the body centered tetragonal PM state (green) and base centered orthorhombic SDW state (black). Projected 2D BZs are shown as shaded planes. (c) Notations shown on 2D projection of respective BZ. We use notations pertaining to the PM BZ (green) throughout, where Γ-X is along the AFM direction, and Γ-Y is along the FM direction. (d) FS in the SDW state measured on twinned crystals in which the X and Y points are mixed and indistinguishable due to existence of twin domains. (e)-(f) FS in the SDW state measured on detwinned crystal along Γ-X and Γ-Y respectively, showing strong anisotropy along these two directions. (g)-(i) Corresponding spectral images along high symmetry lines (top) and their second derivatives (bottom) for the FSs shown above demonstrating the anisotropy in the band dispersions. All data were measured at 10K, with 25eV photons, and polarization vectors labeled in red. All FS presented in the paper are made with an integration window of 5meV about E$_F$.

**Fig. 2: Multi-orbital SDW FS of BaFe$_2$As$_2$ probed by different polarizations**

(a)-(c) FSs probed with different polarizations, indicated by red arrows. Insets (cyan and magenta) show corresponding dotted regions for better contrast. Intensity enhancement of the Y region (cyan inset in (a)) reveals similar FS topology as X region with C$_4$ rotational symmetry broken. All other insets are second derivatives of the corresponding FSs. Different parts of the FS are enhanced or suppressed under different polarizations, indicating multi-orbital character of the FS in the SDW state. Outline of FS visible under each polarization is overlaid on measured

FS in the corresponding panels. (d) A summary of measured FS outlines in the reconstructed SDW BZ. (e) Spectral image cuts through the bright spots along Γ-X showing band crossing near $E_F$. Cuts taken near the Y point in the experimental setup of (a). (f) Spectral image cuts through the petal along Γ-Y showing similar band crossing further below $E_F$. Cuts taken near the Γ point in the experimental setup of Fig. 1e (c3) and Fig. 1f (c4). Electron dispersion is suppressed along the high symmetry cut in c4 due to photoemission matrix elements for this geometry.

**Fig. 3: Temperature and doping dependence: anisotropic band shift observed above $T_{SDW}$ on detwinned Ba(Fe$_{1-x}$Co$_x$)$_2$As$_2$**

(a) Temperature dependence of anisotropic band dispersion (second derivative) along Γ-X and Γ-Y high symmetry lines taken on BaFe$_2$As$_2$ through $T_S/T_{SDW}$ (138K), with 47.5eV photons ($k_z = 0$). Γ-X (Γ-Y) cuts were measured under polarization setup of Fig. 2a (Fig. 2b). Along Γ-X, only $d_{yz}$ orbital is highlighted by photoemission matrix elements on the high symmetry cut, while the same is true for $d_{xz}$ orbital along Γ-Y. Band guides to the eye are drawn to indicate the shift with temperature. Dotted lines are bands from the PM state for comparison. (b) FS of BaFe$_2$As$_2$ taken above $T_S/T_{SDW}$ confirming $C_4$ symmetry, with 54eV photons ($k_z = \pi/2$). The image shown at the X/Y point to the right of Γ is the second derivative of corresponding FS at the X/Y point to the top of Γ for better contrast. (c) Temperature dependence of the anisotropic band dispersion between Γ-X and Γ-Y (second derivatives) on Ba(Fe$_{0.975}$Co$_{0.025}$)$_2$As$_2$ through $T_{SDW}$ (94.5K), with 62eV photons ($k_z = \pi$). (d) Same measurements on unstressed twinned sample for comparison. (e) Energy position of hole-dispersions near X (green) and Y (red) relative to $E_F$ at the representative momentum marked by yellow lines in (c) plotted as a function of temperature, measured on both detwinned and unstressed crystals, compared with resistivity measurements reproduced from Ref.

15. (f) Doping dependence of the $d_{xz}/d_{yz}$ band splitting measured at momentum k=0.9π/a taken at 10K, with 47.5eV photons ($k_z = 0$).

**Fig. 4: Schematic: development of anisotropy in electronic structure of underdoped Ba(Fe$_{1-x}$Co$_x$)$_2$As$_2$**

In the tetragonal PM state (left), both lattice and spin have $C_4$ rotational symmetry which is reflected in the observed $C_4$ symmetric electronic structure. $C_4$ rotational symmetry is also preserved in the orbital content as $d_{xz}$ band along Γ-Y and $d_{yz}$ band along Γ-X are degenerate (solid lines in middle panel). As the system approaches $T_{SDW}$, $C_4$ rotational symmetry in orbital degree of freedom is broken as the $d_{xz}$ band shifts down and $d_{yz}$ band shifts up (dashed lines in middle panel). The anisotropy in the band dispersions is then manifested in the lowered symmetry of the orthorhombic SDW state, as the observed FS is strongly anisotropic reflecting the $C_2$ rotational symmetry of the spin and lattice structure (right). Red (blue) pockets indicate hole (electron) character, whereas magenta indicates pockets resulting from hybridization of electron and hole features.

# Figure 1

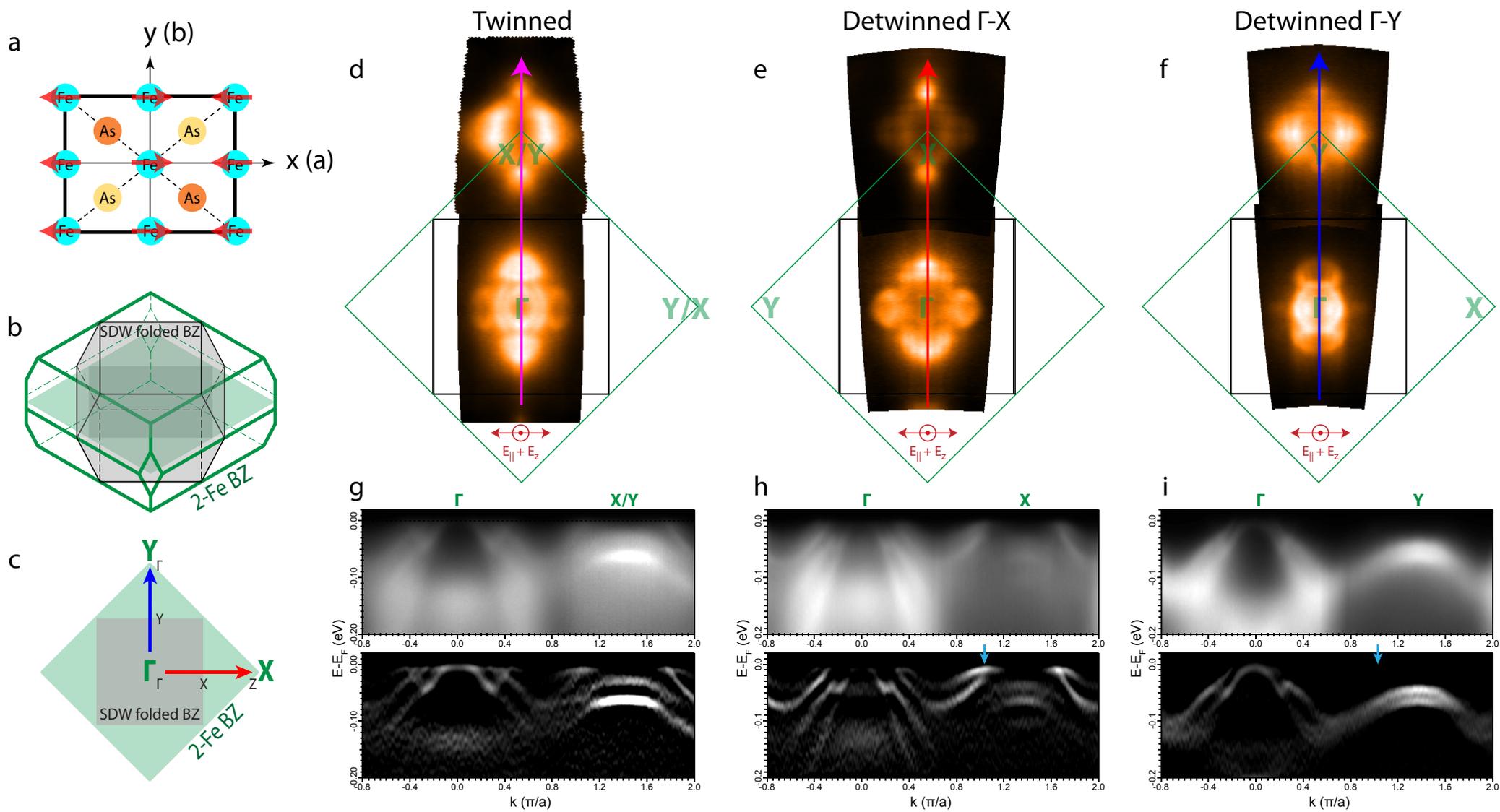

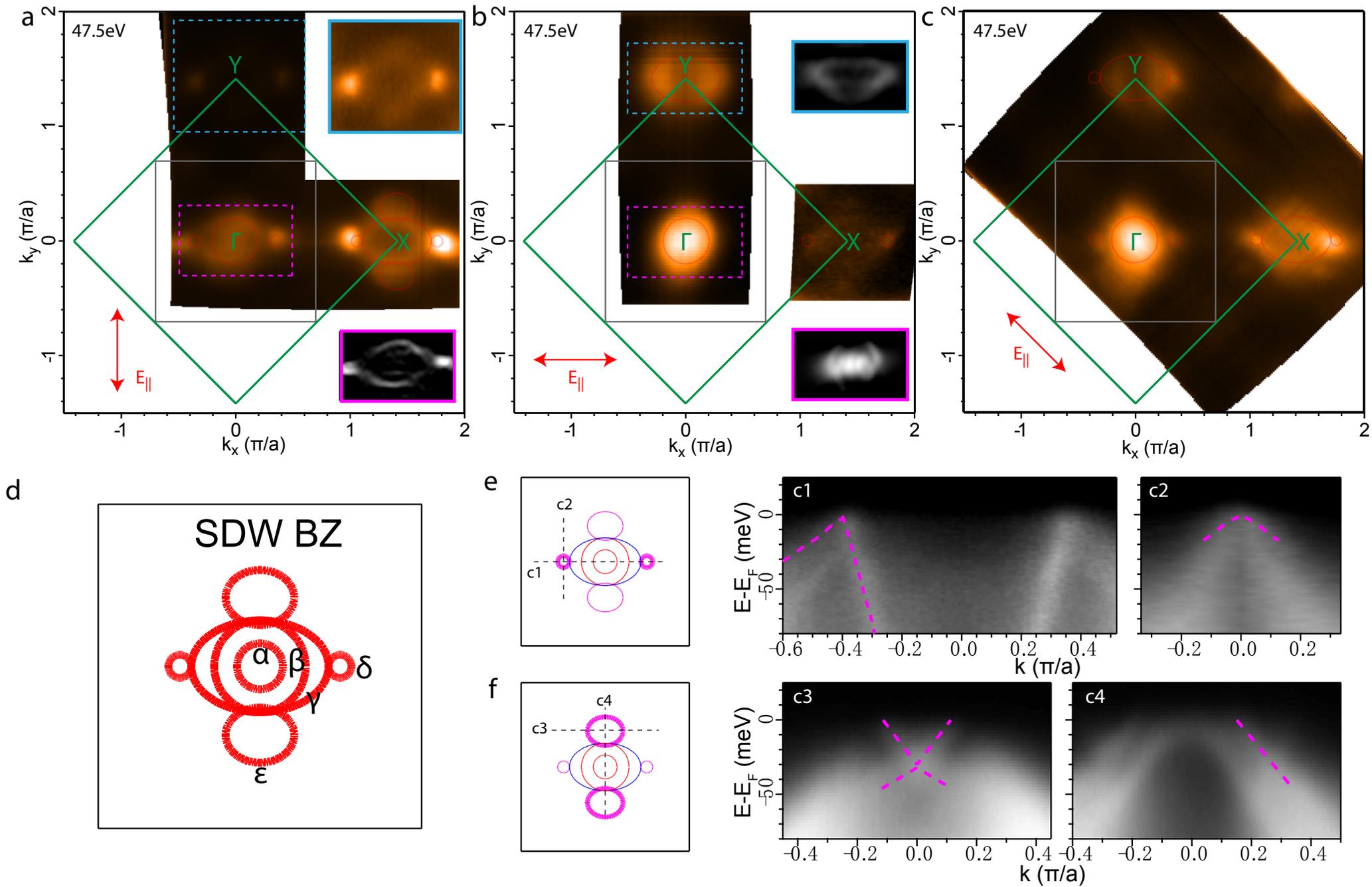

# Figure 3

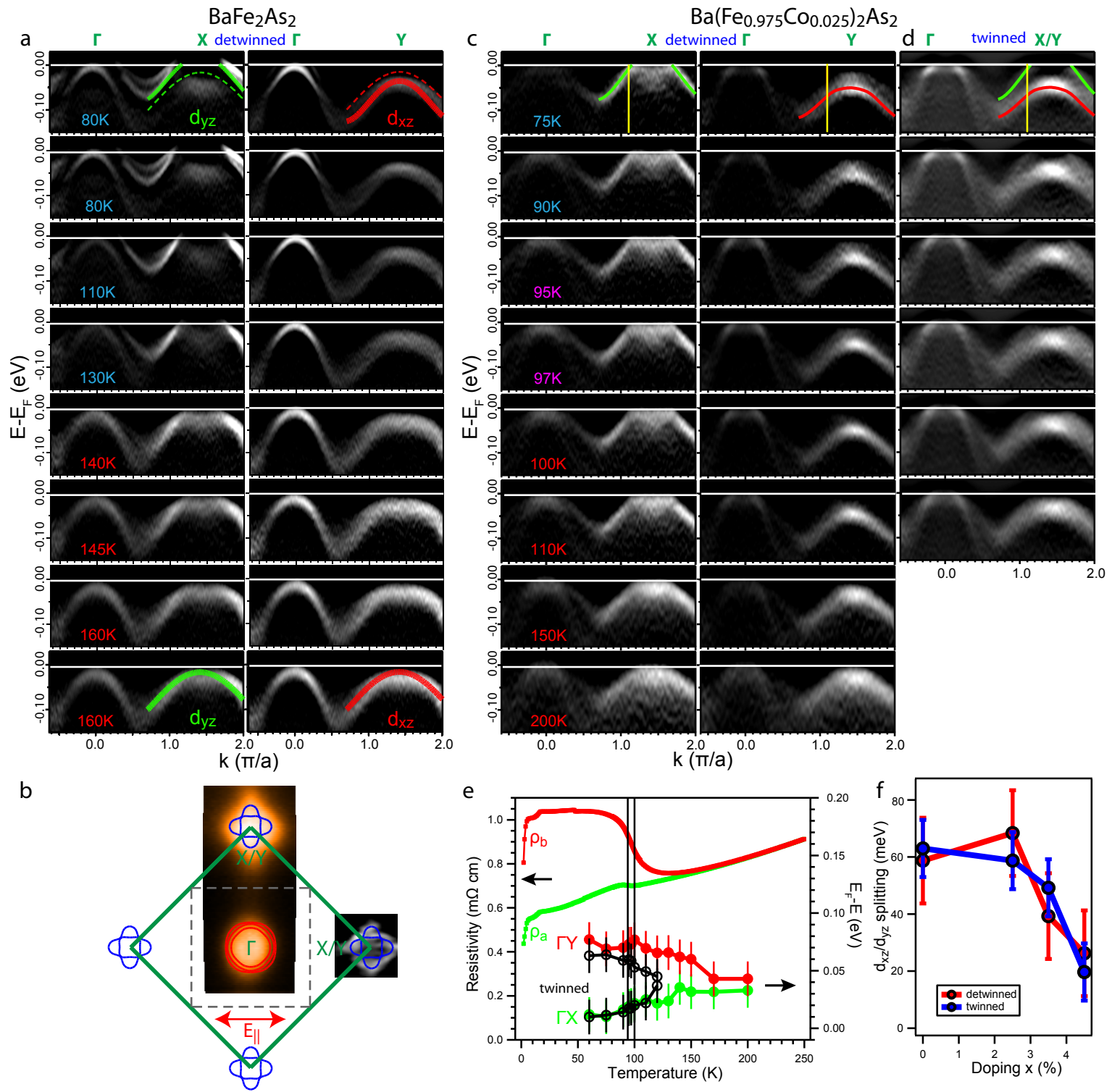

# Figure 4

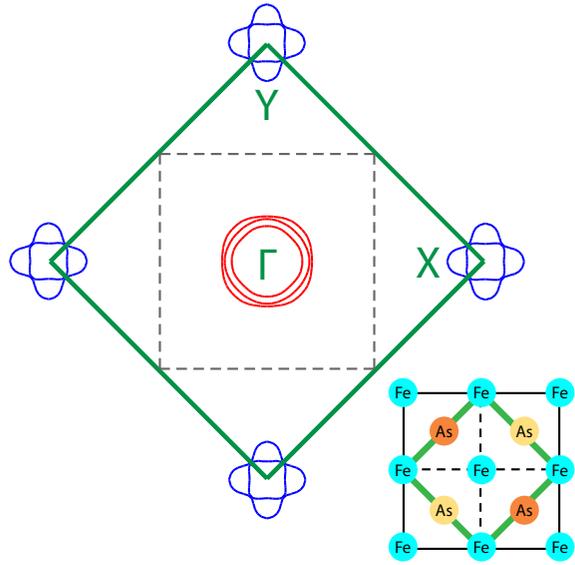
Tetragonal/Paramagnetic

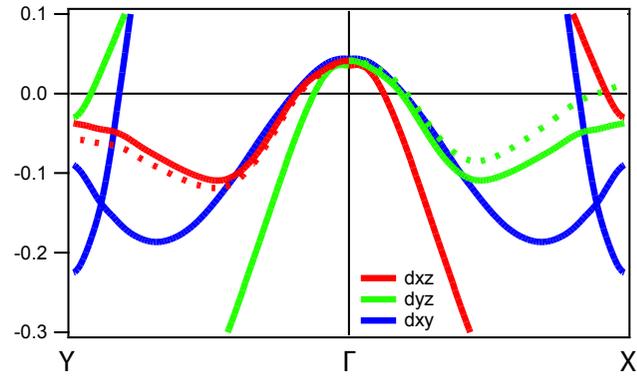
Orbital-Dependent Symmetry Breaking

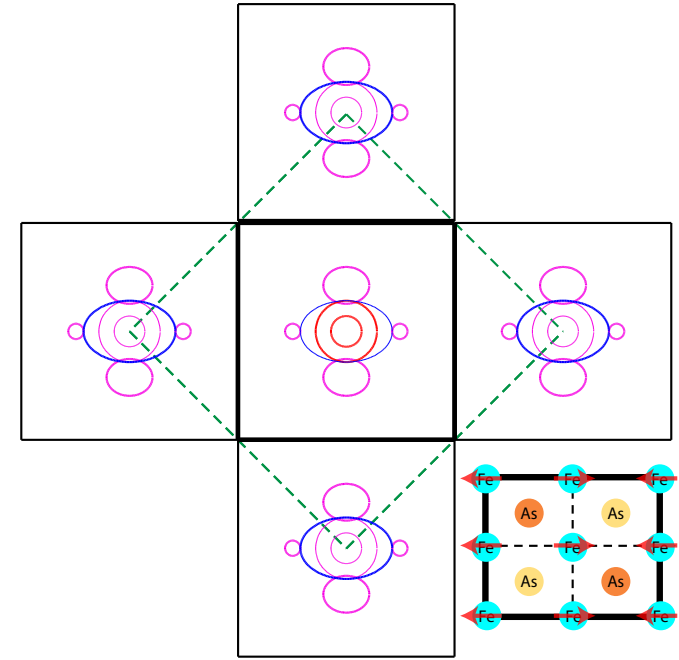
Orthorhombic/SDW

# Symmetry breaking orbital anisotropy observed for detwinned Ba(Fe$_{1-x}$Co$_x$)$_2$As$_2$ above the spin density wave transition

M. Yi, D. H. Lu, J.-H. Chu, J. G. Analytis, A. P. Sorini, A. F. Kemper, B. Moritz, S.-K. Mo, R. G. Moore, M. Hashimoto, W. S. Lee, Z. Hussain, T. P. Devereaux, I. R. Fisher, and Z.-X. Shen

## (Supplementary Information)

**SI.I: Study of k$_z$ dispersion effect on anisotropy of Fermi surface topology**

**SI.II: Discussion of photoemission matrix elements for polarization studies**

**SI. III: Anisotropy in electronic structure observed on Ba(Fe$_{0.975}$Co$_{0.025}$)$_2$As$_2$**

**SI. IV: Analysis details of energy splitting temperature dependence**

**SI. V: Doping dependence of d$_{xz}$ and d$_{yz}$ band splitting**

**SI. VI: Band calculation of lattice distortion effect**

**SI. VII: Model study of orbital anisotropy**

**SI.I: Study of $k_z$ dispersion effect on anisotropy of Fermi surface topology**

In angle-resolved photoemission spectroscopy (ARPES), the electronic structure at different $k_z$ values of the Brillouin zone (BZ) can be probed by varying the energy of the incoming photons. We performed such a study to check the anisotropy of the electronic structure on detwinned crystals of $BaFe_2As_2$ against $k_z$ dispersion. In Fig. S1, pairs of Fermi surface (FS) mapping along Γ-X and Γ-Y are shown for three chosen $k_z$ values: 0 (47.5eV), π/2 (54eV), and π (62eV), covering half period along the $k_z$ direction in a BZ that is symmetric about $k_z$=0 and π. Among these three representative $k_z$ values, we do not observe any qualitative difference in the anisotropy of the FS topology. In particular, the tiny pockets observed as bright spots along $k_x$ and the bigger petal pockets along $k_y$ show little changes for all $k_z$ values. Moreover, in Fig. S1g and S1h we show a more detailed $k_z$ dispersion along both Γ-X and Γ-Y directions. We note that the highly dispersive feature apparent around the Γ point in Fig. S1h is the inner hole pocket α, which closes near $k_z$ = 0 and opens the widest at $k_z$ = π. All the other features, including those associated with the most anisotropic features of bright spots and petals remain mostly 2D. Hence, the in-plane anisotropy of the FS topology is robust even when the $k_z$ dispersion is taken into account.

**SI.II: Discussion of photoemission matrix elements for polarization studies**

ARPES directly probes the one-particle spectral function A(**k**, ω). However, the measured ARPES intensity is also affected by the matrix elements[1]. For the multi-band, multi-orbital iron pnictides, the polarization matrix elements play an important role in ARPES studies. In simple terms, the ejected photoelectrons from each orbital are subject to symmetry selection rules based on the orientation of the polarization of the incoming photons as well as the orbital symmetry with respect to the crystal axes, $I_0 \propto |\langle \phi_f^{\mathbf{k}} | \boldsymbol{\varepsilon} \cdot \mathbf{x} | \phi_i^{\mathbf{k}} \rangle|^2$, where $\boldsymbol{\varepsilon}$ is a unit vector along the polarization direction of the incoming photons. For illustration purposes, consider the high symmetry cut along the $x$ direction which defines a mirror plane as shown in Fig. S2. As the final state $\phi_f^{\mathbf{k}}$ is even with respect to this mirror plane, for a polarization that is odd with respect to this mirror plane ($E_s$ in this case and also the case for Γ-X cut in Fig. 2a), only those orbitals ($d_{yz}$, $d_{xy}$) in the initial state $\phi_i^{\mathbf{k}}$ with odd parity with respect to this mirror plane will have non-zero contribution to the measured signal, while those with even parity ($d_{z2}$, $d_{x2-y2}$, $d_{xz}$) will be suppressed along this high symmetry cut. For Γ-Y cut in Fig. 2b, the parity between $d_{xz}$ and $d_{yz}$ switches while the rest remain the same. Therefore, for a multi-orbital system, different polarization geometries would highlight or suppress different orbitals according to their symmetries. This effect is especially important for the iron pnictides because nearly all five Fe 3d orbitals are active near $E_F$. Hence, full polarization studies are important to observe the complete electronic structure.

In this regard, the information contained in the enhancement or suppression of different bands and/or associated FS sheets in certain polarization geometries can be utilized in understanding the contribution of different orbital characters to the electronic structure. For this purpose, we have calculated the polarization matrix elements (Fig. S2b-c) for different orbitals under two polarization geometries corresponding to those used for the Γ-X and Γ-Y high symmetry cuts in Fig. 3a and 3c. Now we discuss the identification of the $d_{xz}$ and $d_{yz}$ bands for the temperature dependence studies. For all the cuts in Fig. 3a and 3c, the polarization used was perpendicular to each cut direction, meaning that the matrix elements in Fig. S2(b) match those of the Γ-X cuts while those of Fig. S2(c) match those of the Γ-Y cuts. Hence for the Γ-X cuts, only bands of $d_{yz}$ and $d_{xy}$ orbitals have intensity along this high symmetry cut. However, as shown in the calculations in Fig. S2(b), the $d_{xy}$ orbital has very weak intensity in the first BZ where the cut was measured. Hence what we observe along the Γ-X cut in Fig. 3 must be of dominant $d_{yz}$ character. The same is true of the observation of $d_{xz}$ band along Γ-Y. Therefore, the observation of the anisotropic shifting of bands with temperature is a manifestation of anisotropy in the orbital degree of freedom.

## SI. III: Anisotropy in electronic structure observed on Ba(Fe$_{0.975}$Co$_{0.025}$)$_2$As$_2$

Measurements on detwinned doped Ba(Fe$_{0.975}$Co$_{0.025}$)$_2$As$_2$ reveals very similar anisotropy in the electronic structure as that of undoped. We again observe on the FS, bright spots only reside along the Γ-X direction (Fig. S3a) while petal-like features reside along the Γ-Y direction (Fig. S3b), breaking the C$_4$ rotational symmetry in the same way as that of undoped. These features are again the results of band crossings at different energy positions (Fig. S3e-f), leading to Fermi pockets of different sizes on the two orthogonal directions. The high symmetry cuts along these two directions also exhibit strong anisotropic band shifts, with the hole-like dispersion near the X point raised across the E$_F$, while that near the Y point is shifted down in the orthorhombic SDW state (Fig. S3c-d).

## SI. IV: Analysis details of energy splitting temperature dependence

In this section we provide the analysis details that led to the temperature dependence trace of the $d_{yz}$ and $d_{xz}$ band separation shown in Fig. 3e in the main text, which is measured on Ba(Fe$_{0.975}$Co$_{0.025}$)$_2$As$_2$. In Fig. S4a, we show the spectral images taken at the lowest temperature of the series shown in Fig. 3c-d. At this temperature (60K), which is well below $T_{SDW}$ (94.5K), we can see that the $d_{yz}$ band along Γ-X and the $d_{xz}$ band along Γ-Y are shifted in the opposite direction, as measured on detwinned crystals. These orthogonal bands can also be measured simultaneously on twinned crystals as they appear on the same cut due to domain mixing (Fig. S4a). To see how these bands shift with temperature, we take a representative momentum (marked by yellow line in Fig. (S4a)), and trace the energy positions of the relevant bands with temperature. The energy position of a band is identified by a peak in the energy distribution curve (EDC) (Fig. S4b). This is easily identifiable if the relevant band is sufficiently far away from other bands, as is the case for detwinned crystals, but less clear for the twinned crystal, where the domain-mixed $d_{yz}$ and $d_{xz}$ bands are close together in energy. To better trace these features, we take the second derivative of the EDCs, where local minima correspond to the peak positions in the raw EDCs. We compare the raw EDCs and their respective second derivatives in Fig. S4b-c for the momentum marked in yellow, where the energy positions of the $d_{yz}$ and $d_{xz}$ bands can be identified by the two well-separated minima in the second derivatives measured on both detwinned and twinned crystals. When such analysis is performed for a series of temperatures, we see that the $d_{yz}$ band indeed shifts toward the Fermi level, whereas the $d_{xz}$ band shifts away from the Fermi level as temperature is lowered (Fig. S4d-f). The minima marked by diamonds in Fig. S4d-f are the energy positions used to plot Fig. 3e in the main text.

**SI. V: Doping dependence of $d_{xz}$ and $d_{yz}$ band splitting**

In Fig. S5a we compare the second derivative of the spectral images taken along the Γ-X and Γ-Y high symmetry directions for four different dopings across the underdoped region. We see clearly that the $d_{yz}$ and $d_{xz}$ bands are anisotropically shifted along these orthogonal directions. We also compare side by side the same measurements taken on twinned crystals where the $d_{yz}$ band from Γ-X direction and $d_{xz}$ band from Γ-Y direction are observed on the same cut due to domain mixing. To compare the magnitude of the anisotropic band shift across doping, we again take a representative momentum (marked by yellow line) along the hole-like dispersions, and measure the energy positions of the $d_{yz}$ and $d_{xz}$ bands as indicated by the minima of the second derivative of the EDC. In Fig. S4c, we mark the energy positions of the $d_{yz}$ and $d_{xz}$ bands for both detwinned and twinned crystals. Firstly, we make the observation that at this base temperature (10K) well below $T_{SDW}$, the detwinning stress does not affect the intrinsic magnitude of the band splitting as the $d_{xz}$ and $d_{yz}$ bands from detwinned crystals are observed to be at the same positions as those measured on unstressed twinned crystals. Furthermore, we see that the separation between the $d_{xz}$ and $d_{yz}$ bands decreases monotonically with doping, as this energy separation is plotted in Fig. 3f in the main text.

**SI. VI: Band calculation of lattice distortion effect**

Here we present the effect of the observed lattice distortion alone on the band structure. Non-magnetic calculations were performed using WIEN2k[2], using the experimentally observed low temperature (10K) orthorhombic lattice distortion reported by Ni *et al*[3]. Calculated bands are renormalized by a factor of 2 to provide a more appropriate match to the renormalized band dispersions observed by ARPES (Fig. S6). The splitting between the relevant $d_{yz}$ (green) and $d_{xz}$

(red) bands due to this orthorhombic distortion alone is calculated to be on the order of 10meV, which is too small to explain the observed value of 80meV. Hence the observed electronic anisotropy cannot solely result from the lattice distortion in the orthorhombic state below $T_S$.

**SI. VII: Model study of orbital anisotropy**

Here we present the effect of the orbital degeneracy splitting on the band structure and orbital occupations as calculated within a tight-binding model. We use the model of Graser *et al.*[4], which is an appropriate description of the three-dimensional band structure of $BaFe_2As_2$. The model as is, however, does not exactly match the experimentally observed bands; this slight mismatch is not uncommon for models based on LDA calculations as previously reported[5] and is likely to be a result of strong interband scattering process[6,7]. We find that to match the experimental band structure, one needs to apply an upward shift of 0.2eV near the X point, and an overall renormalization of the band structure by a factor of 3. Due to limits in experimental condition, the shift near Gamma is less precise. However, the quantities we will present here do not strongly depend on the shift of the band structure, but rather on the magnitude of orbital splitting. As such, the exact method of shifting the band structure does not strongly bear on these results.

Figure S7 shows the band structure and orbital character of the bands, calculated from the tight-binding model after application of a splitting, $\Delta$, between the on-site energies of the $d_{xz}$ and $d_{yz}$ orbitals. As illustrated by the figure, with no splitting ($\Delta = 0$), the band structure is $C_4$ symmetric; the $d_{yz}$ bands are equivalent to the $d_{xz}$ bands after a rotation by $\pi/2$. Upon application of a 60meV shift that matches the experimental value, the $d_{xz}$ and $d_{yz}$ bands appear markedly different. The $d_{yz}$ bands are raised (by 45meV), and the $d_{xz}$ bands are lowered (by 15meV). Since the simple on-site orbital energy shift produces a band structure that is quite similar to the experimentally observed one, we can extract a measure of the anisotropy in orbital occupations from the model. With a 60 meV splitting, we obtain an estimated 18% difference in orbital occupations. We would like to point out that this number depends, albeit weakly, upon the exact method and amount of band shift applied to ensure the agreement of the model with the ARPES data as a starting point. It does, however, depend linearly upon the orbital energy shift. Nevertheless, the estimated difference in orbital occupation of 18%, with an error bar of a few percent, can be used as a guide for calculations of the effects of orbital polarization, e.g. as done by Chen *et al.*[8]

**Figure Legends**

**Fig. S1: Study of k$_z$ dispersion on detwinned samples**

(a)-(c) Fermi surface mapping on detwinned BaFe$_2$As$_2$ along the Γ-X direction at 47.5eV, 54eV, and 62eV, corresponding to k$_z$ values of 0, π/2, and π respectively, at the Γ point in the BZ notation with a period of 2π/c along the k$_z$ direction. The photon energies of 47.5eV and 62eV represent the largest difference in k$_z$ dispersion at the Γ point due to its symmetry about k$_z$=0 and π. (d)-(f) Same as (a)-(c) but taken along the Γ-Y direction. Polarization vectors are all in-plane and labeled in red. (g) Detailed k$_z$ dispersion mapping along the Γ-X direction measured under the same polarization as that in (a)-(c). (h) k$_z$ dispersion along Γ-Y under the same polarization as that in (d)-(f). All measurements were performed at 10K.

**Fig. S2: Polarization matrix elements**

(a) Schematic showing the relation of polarization vectors ($E_s$, $E_p$) to the crystal axes. (b) Matrix element intensity calculated for the five 3d orbitals for the experimental geometry in which the polarization of the incoming photons is in-plane along the Γ-Y direction. Lighter color indicates higher intensity. (c) Same as (b) but with polarization vector rotated 90° along Γ-X direction.

**Fig. S3: Anisotropy in electronic structure of Ba(Fe$_{0.975}$Co$_{0.025}$)$_2$As$_2$**

(a)-(b) Fermi surface mapping on detwinned Ba(Fe$_{0.975}$Co$_{0.025}$)$_2$As$_2$ along the Γ-X and Γ-Y directions, respectively, with 62eV photons ($k_z = \pi$). Polarization vectors are labeled in red. (c)-(d) High symmetry cuts along Γ-X and Γ-Y directions showing the anisotropy of the band dispersions, taken with 62eV photons. (e)-(f) Same cuts as those for undoped in Fig. 3e-f, showing the anisotropic band crossing in doped compound that result in the bright spots on the FS along Γ-X and petal-like features along Γ-Y. All measurements were performed at 10K.

**Fig. S4: Temperature dependence of energy splitting in Ba(Fe$_{0.975}$Co$_{0.025}$)$_2$As$_2$**

(a) Second derivatives of spectral images taken along the Γ-X and Γ-Y high symmetry directions on detwinned crysta and twinned crystal, taken at 60K with 62eV photons. (b) EDCs at the momentum marked by yellow line of the raw spectral images corresponding to the spectral images shown in (a). (c) Second derivatives of the EDCs shown in (b). The curve for twinned crystal is offset for clarity. (d)-(e) Second derivatives of the EDCs at the yellow line in (a) as a function of temperature, taken along the Γ-X and Γ-Y high symmetry directions of detwinned crystals, respectively. (f) Same measurements taken on twinned crystal. The diamonds indicate the dips in the second derivative curves where peaks are expected in the raw EDCs, which are the energy positions of the bands at this momentum.

**Fig. S5: Doping dependence of orbital anisotropy**

(a) Second derivatives of spectral images taken along the Γ-X and Γ-Y high symmetry directions on detwinned crystals of four different dopings. (b) Same measurement on twinned crystals. The green (red) lines are guide to eye of the $d_{yz}$ ($d_{xz}$) band. Solid lines are all drawn at the same energy position for comparison, whereas dotted lines show the shifted position of the bands for 4.5% doping. (c) Second derivatives of the EDCs taken at the representative momentum marked by yellow lines in (a)-(b). The minima marked by the arrows are the energy positions of the corresponding $d_{yz}$ and $d_{xz}$ band on detwinned and twinned crystals. All measurements were taken at 10K with 47.5 eV photons ($k_z$=0). The dotted lines for twinned crystals are offset for clarity.

**Fig. S6: Effect of lattice distortion on band structure**

Band structure along high symmetry directions obtained from non-magnetic calculations using WIEN2k accounting for the orthorhombic lattice distortion. Bands renormalized by a factor of 2 to better match the ARPES energy scales. The orbital characters of the bands are marked in color.

**Fig. S7: Effect of orbital shift on band structure**

Band structure for Δ=0 (left) and Δ=60meV (right). The fraction of $d_{xz}$ (bottom) and $d_{yz}$ (top) orbital composition is indicated by color and linewidth.

Figure S1

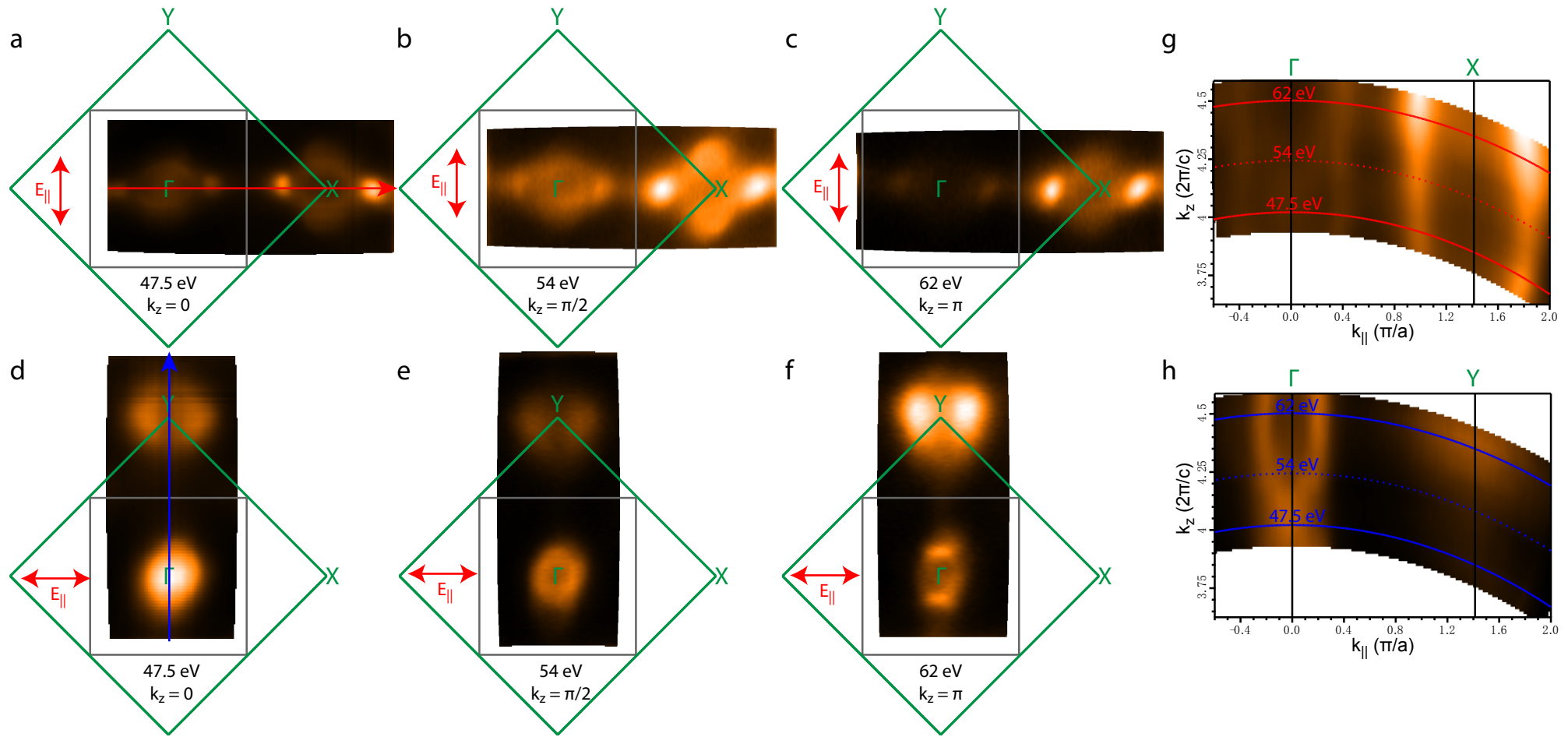

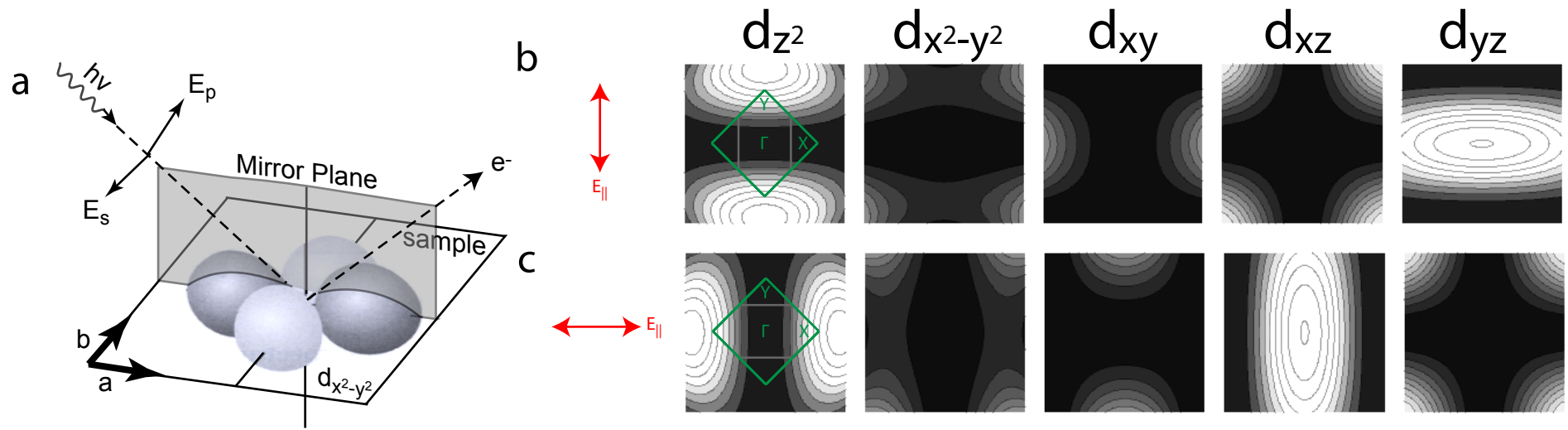

# Figure S3

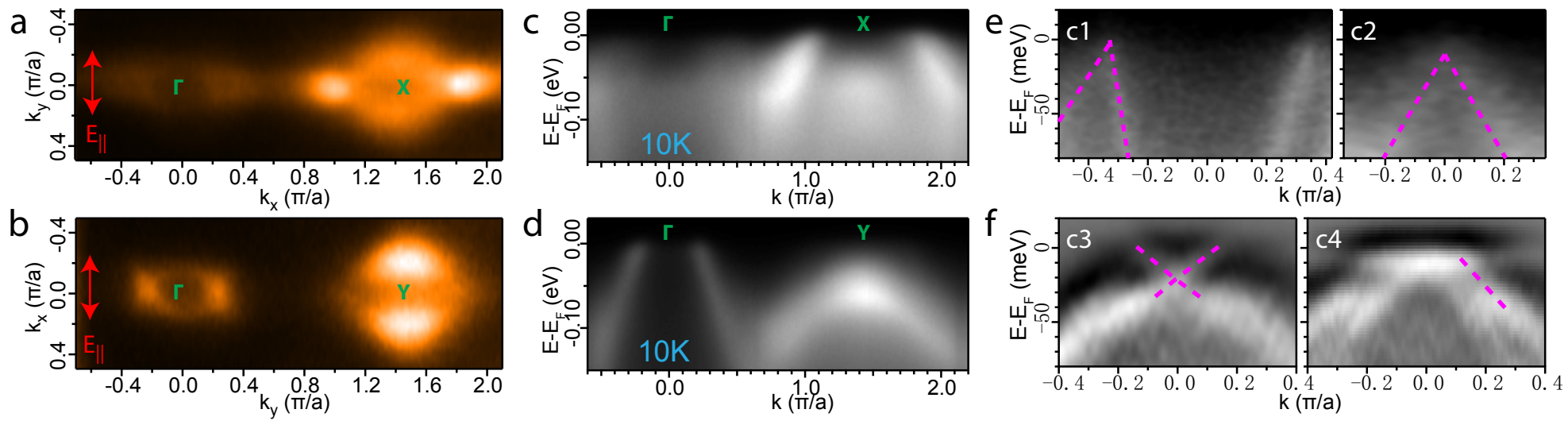

Figure S4

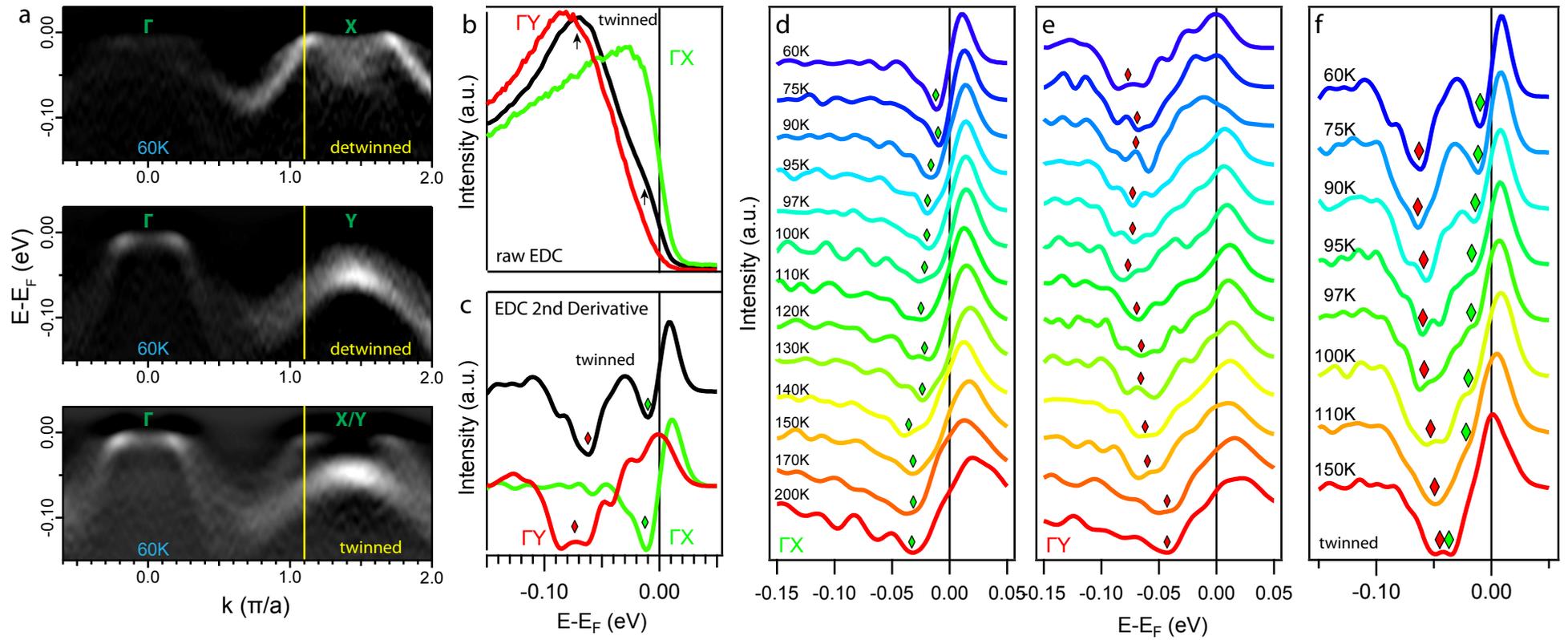

# Figure S5

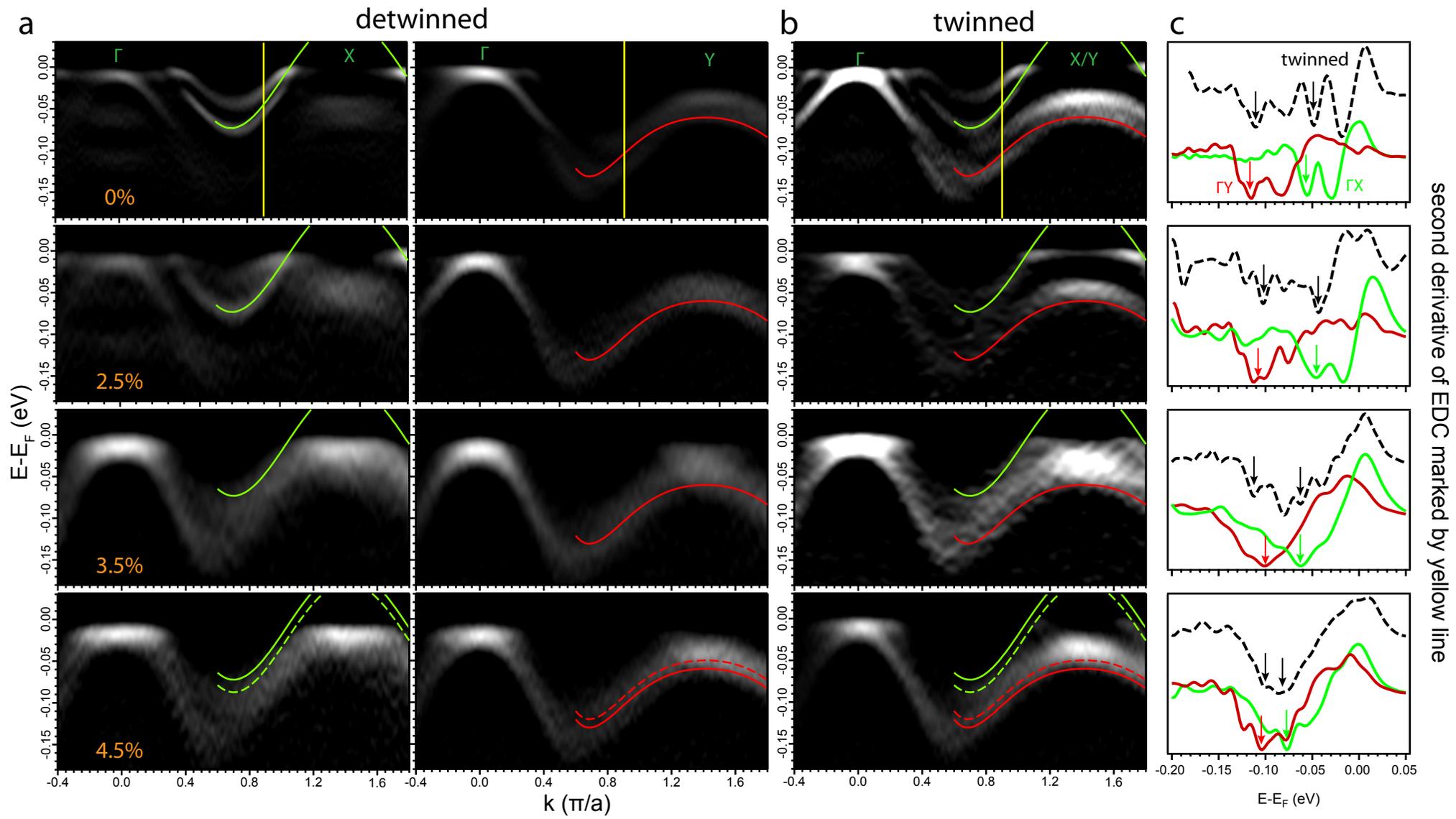

Figure S6

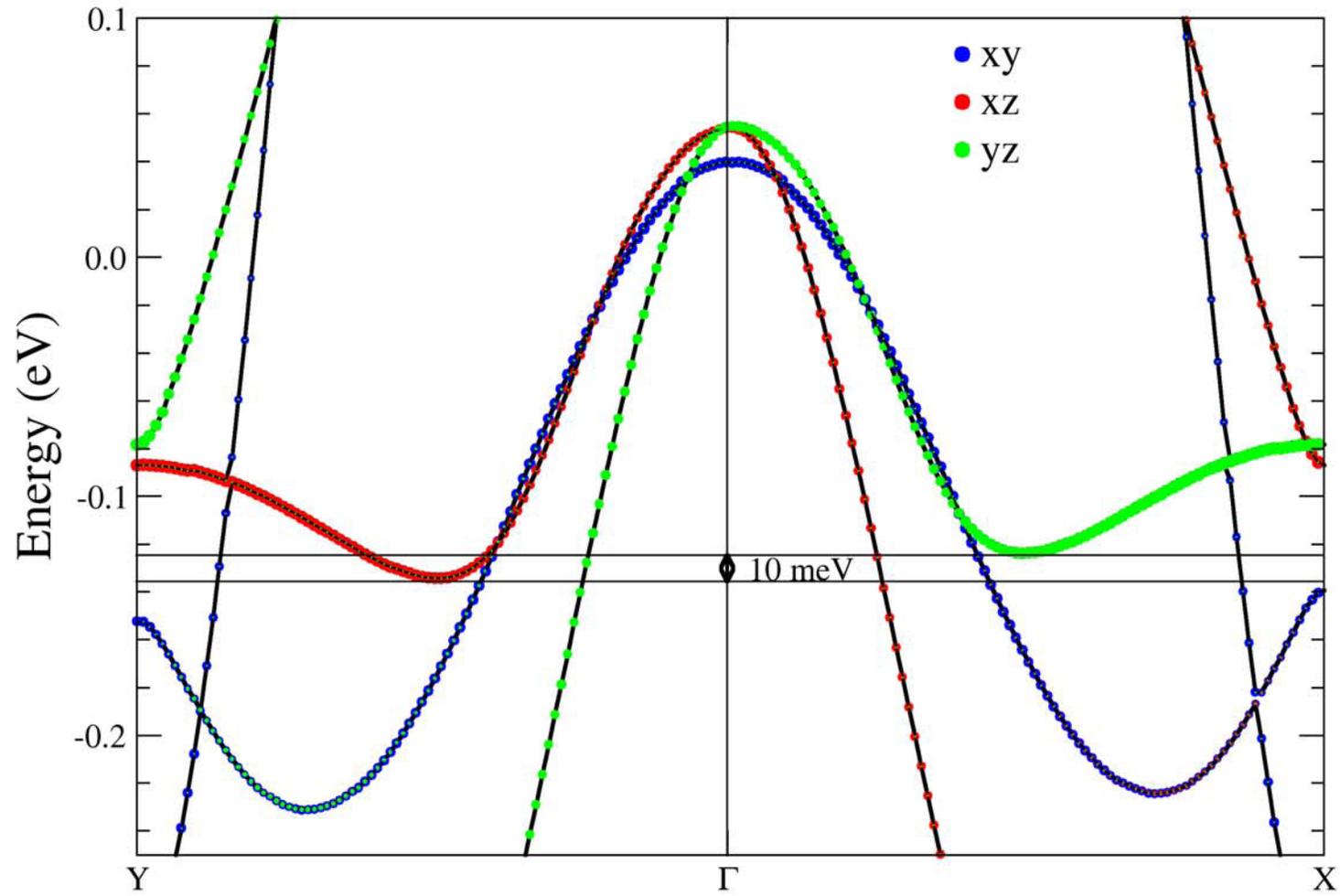

# Figure S7

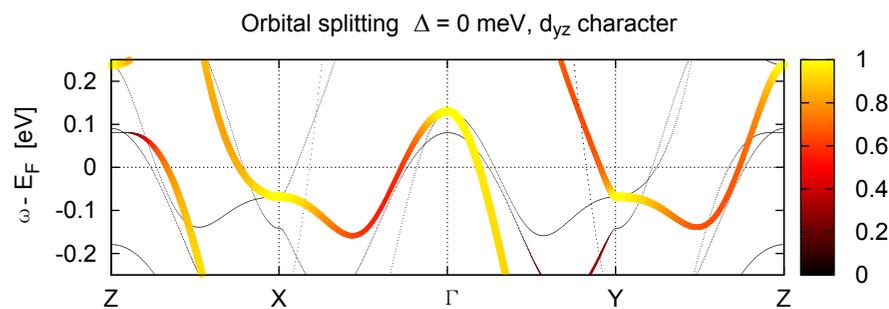
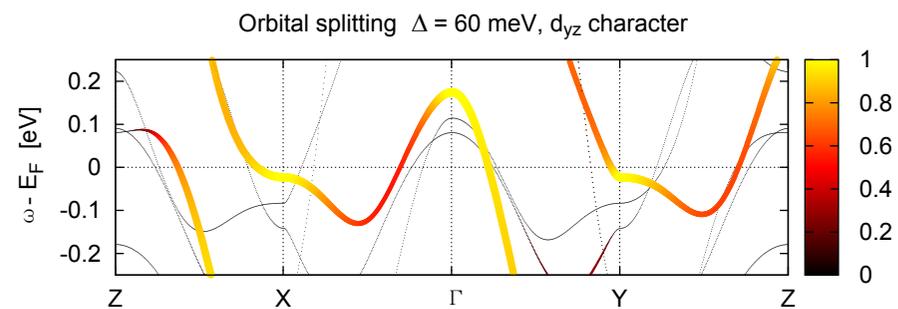
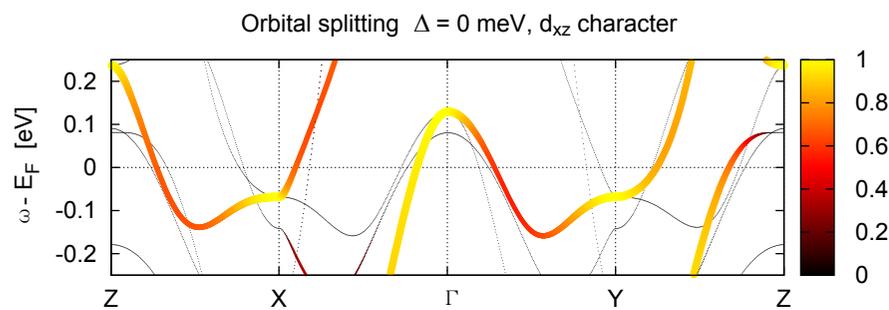
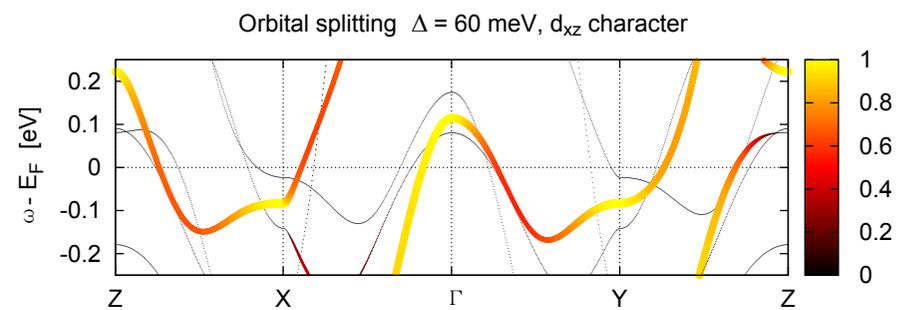